\documentclass[11pt]{article}
\usepackage{fullpage,epsf,amssymb,amsthm,amsfonts,amsmath,latexsym,array,graphicx,extarrows,mathtools,mathstyle,mathrsfs,slashed,subcaption,amsbsy,csquotes,accents}
\usepackage[normalem]{ulem}
\usepackage{authblk}
\usepackage{cite}
\usepackage{color}

\usepackage{float}

\usepackage{authblk}

\usepackage{hyperref}

\usepackage[export]{adjustbox}

\numberwithin{equation}{section}

\newcommand{\overbar}[1]{\mkern 1.5mu\overline{\mkern-1.5mu#1\mkern-1.5mu}\mkern 1.5mu}

\title{\bf{Non-perturbative BRST symmetry and the spectral structure of the ghost propagator}}

\author[1]{Shirley Weishi Li\thanks{shirleyl@fnal.gov}}
\author[2]{Peter Lowdon\thanks{lowdon@itp.uni-frankfurt.de}}
\author[3]{Orlando Oliveira\thanks{orlando@uc.pt}}
\author[3]{Paulo J.~Silva\thanks{psilva@uc.pt}}

\affil[1]{Theoretical Physics Department, Fermi National Accelerator Laboratory, PO Box 500, Batavia, IL 60510, USA}
\affil[2]{Institut f\"{u}r Theoretische Physik, Johann Wolfgang Goethe-Universit\"{a}t, Max-von-Laue-Str. 1,  60438 Frankfurt am Main, Germany}
\affil[3]{Centro de F\'{i}sica da Universidade de Coimbra, Departamento de F\'{i}sica, Universidade de Coimbra, 3004-516 Coimbra, Portugal}

\date{}
\begin{document}
\begin{flushright}   \end{flushright}
\vspace{15mm} 
{\let\newpage\relax\maketitle}
\setcounter{page}{1}
\pagestyle{plain}

\begin{abstract}
\noindent
In BRST-quantised Yang-Mills theory the existence of BRST symmetry imposes significant constraints on the analytic structure of the continuum theory. In particular, the presence of this symmetry in the non-perturbative regime implies that any on-shell state with vanishing norm must have an associated partner state with identical mass, but negative inner product. In this work we demonstrate that for quantum chromodynamics (QCD) this constraint gives rise to an interconnection between the ghost and gluon spectra, and in doing so provides a non-trivial test of whether BRST symmetry is realised non-perturbatively. By analysing infrared lattice data for the minimal Landau gauge ghost propagator in pure $\mathrm{SU}(3)$ Yang-Mills theory, and comparing this with previous results for the gluon propagator, we show that this interconnection is violated, and hence conclude that continuum and current lattice formulations of Yang-Mills theory in Landau gauge represent two distinct realisations of the theory.
\end{abstract}

\newpage

\section{Introduction}
\label{intro}

Understanding the characteristics of correlation functions involving the fundamental fields in Yang-Mills theory is essential for unravelling the non-perturbative structure of the theory. In the local Becchi-Rouet-Stora-Tyutin (BRST) quantisation of quantum chromodynamics (QCD) the ghost degrees of freedom play a central role, with the ghost propagator in particular having important phenomenological consequences such as in the determination of glueball masses~\cite{Huber:2020ngt}, as well as potential implications for the nature of confinement itself~\cite{Kugo:1979gm,Alkofer:2000wg,Alkofer:2006fu}. The calculation of the ghost propagator has been a research focus for many years, resulting in numerous studies based on a variety of different non-perturbative calculational approaches, including functional methods~\cite{Alkofer:2000wg,vonSmekal:1997ohs,Aguilar:2008xm,Boucaud:2008ky,Fischer:2008uz,Horak:2021pfr,Aguilar:2021okw} and lattice Monte Carlo simulations~\cite{Suman:1995zg,Oliveira:2006yw,Oliveira:2007dy,Cucchieri:2007md,Cucchieri:2008fc,Sternbeck:2012mf,Duarte:2016iko,Boucaud:2018xup,Maas:2019ggf}. In recent years, advanced numerical inversion techniques have also been applied to the subsequent data from these studies in order to gain new insights into the structure of the corresponding ghost spectral density~\cite{Dudal:2019gvn,Binosi:2019ecz,Horak:2021syv}. Nevertheless, despite this intense activity it still remains an open question as to how the ghost propagator should behave, particularly in the low-momentum \textit{infrared} regime, and what bearing this has on the spectrum of the theory. \\

\noindent
Although the precise structure of correlation functions in BRST-quantised QCD remains far from settled, progress has been made by making use of analytic non-perturbative constraints imposed on the theory by the structural assumptions of local quantum field theory (QFT)~\cite{Streater:1989vi,Haag:1992hx,Nakanishi:1990qm,Bogolyubov:1990kw,Strocchi13}, including locality and Poincar\'{e} covariance. A constraint of particular importance is the K\"{a}ll\'{e}n-Lehmann representation~\cite{Kallen:1952zz,Lehmann:1954xi}, the existence of which demonstrates that the behaviour of correlation functions is controlled by spectral densities whose singularities reflect the presence of on-shell states in the theory\footnote{Although the focus of this study is on vacuum-state correlation functions, spectral-type representations can also be proven to exist more generally for non-vanishing temperatures~\cite{Lowdon:2021ehf}.}. In local gauge theories such as BRST-quantised QCD the necessary presence of states with a non-positive inner product gives rise to additional subtleties, specifically for correlation functions involving coloured fields~\cite{Lowdon:2015fig,Lowdon:2017uqe,Lowdon:2017gpp,Lowdon:2018mbn,Li:2019hyv}. Since the ghost propagator is the central focus of this work, it is important to understand how these subtleties can potentially affect the behaviour of the corresponding ghost spectral density. This will be discussed in Sec.~\ref{spectral_ghost}. \\
    
\noindent
It is a well-known problem in lattice formulations of Yang-Mills theory that the construction of a well-defined non-perturbative path integral is non trivial due to the infamous Gribov-Singer ambiguity~\cite{Gribov:1977wm,Singer:1978dk}. In the case of BRST-quantised Yang-Mills theory, this ambiguity results in degenerate solutions of the lattice gauge-fixing condition, so-called Gribov copies. The appearance of these copies has been shown to result in numerous complications including the Neuberger problem, which implies that the expectation value of gauge-invariant observables has the ill-defined $0/0$ form~\cite{Neuberger:1986xz}. Several solutions have been proposed to avoid this obstacle, including specific restrictions of the path integral domain~\cite{Zwanziger:1989mf,Vandersickel:2012tz}, topological approaches~\cite{Kalloniatis:2005if}, and the introduction of additional degrees of freedom~\cite{Pelaez:2021tpq}, although no lattice implementation of these methods have been established so far. In most practical lattice calculations the path integral is restricted in such a way that the path integral is well defined, but Gribov copies can still potentially have an effect on the results. An important such example is minimal Landau gauge~\cite{Zwanziger:1991gz,vanBaal:1991zw}. The analysis of ghost propagator lattice data in this gauge forms a central component of this study, and so more details of the gauge will be discussed in Sec.~\ref{spectral_ghost}.    \\

\noindent
The goal of this study is to use a combination of lattice data and analytic arguments in order to gain a better understanding of the infrared characteristics of the ghost propagator, and its effect on the non-perturbative structure of BRST-quantised QCD. The remainder of the paper is organised as follows: in Sec.~\ref{spectral_ghost} we outline the analytic constraints imposed on the ghost propagator by the structural assumptions of local QFT, in Sec.~\ref{brst_spec} we use the assumption of BRST symmetry to establish a non-trivial connection between the infrared properties of the gluon and ghost propagators, in Sec.~\ref{lattice_fits} we analyse lattice data for the ghost propagator in minimal Landau gauge in order to test the condition derived in Sec.~\ref{brst_spec}, and finally in Sec.~\ref{concl} we summarise our findings.

\section{Analytic structure of the ghost propagator}
\label{spectral_ghost}

In BRST-quantised QCD the ghost propagator can be shown~\cite{Lowdon:2017gpp} to have the general non-perturbative form\footnote{Here we adopt a slightly different convention to that used in~\cite{Lowdon:2017gpp} by defining the Fourier transform of the ghost two-point function $\mathcal{F}\left[\langle 0| C^{a}(x)\overbar{C}^{b}(y)|0\rangle \right] =  \mathcal{P}^{ab}_{C}(\partial^{2})\delta(p) + 2\pi \int_{0}^{\infty} ds \, \theta(p^{0})\delta(p^{2}-s) \rho_{C}^{ab}(s)$. This amounts to making the replacement $\rho_{C}^{ab}(s)\rightarrow 2\pi\rho_{C}^{ab}(s)$ in the results of~\cite{Lowdon:2017gpp}, and ultimately removes the $2\pi$ factor appearing in the denominator of the propagator.}
\begin{align}
G^{ab}(p) &=   i\int_{0}^{\infty} \!\! ds \, \frac{ \rho_{C}^{ab}(s) }{p^{2}-s +i\epsilon}  + \mathcal{P}_{C}^{ab}(\partial^{2})\delta^{4}(p),
\label{ghost_prop}
\end{align}
where $\mathcal{P}_{C}^{ab}(\partial^{2})$ is a finite order polynomial in the d'Alembert operator $\partial^{2} = \frac{\partial}{\partial p_{\mu}}\frac{\partial}{\partial p^{\mu}}$. As detailed in~\cite{Lowdon:2017gpp,Lowdon:2017uqe}, the potential non-vanishing of coefficients in $\mathcal{P}_{C}^{ab}(\partial^{2})$, and hence the appearance of $\delta^{4}(p)$ derivatives, stems from the fact that the theory is quantised with an indefinite inner product, and hence necessarily contains states which violate positivity~\cite{Kugo:1979gm}. Since the goal of this work is to use the constraint of Eq.~\eqref{ghost_prop} in order to analyse lattice data for the ghost propagator, we will ignore the possibility of purely singular terms at $p=0$ and focus instead on the first term, which has the standard K\"{a}ll\'{e}n-Lehmann form for a scalar field~\cite{Kallen:1952zz,Lehmann:1954xi}.

\subsection{Ghost spectral density}
\label{ghost_den}

Equation~\eqref{ghost_prop} demonstrates that the analytic form of the ghost propagator is controlled by the properties of the corresponding spectral density $\rho_{C}^{ab}(s)$. Whilst $\rho_{C}^{ab}(s)$ can in principle contain continuous contributions as well as ordinary mass components of the form $\delta(s-m^{2})$, where $m\geq 0$, it was demonstrated in~\cite{Oehme:1979bj,Lowdon:2015fig} that the lack of state-space positivity opens the door for the possibility of a broader class of singular components, so-called \textit{generalised pole} terms
\begin{align}
\delta^{(n)}(s-m_{n}^{2}) = \left(\tfrac{d}{ds}\right)^{n}\delta(s-m_{n}^{2}), \quad\quad  n \geq 1.
\label{gen_pole}
\end{align}
It turns out that the terms in Eq.~\eqref{gen_pole}, together with standard mass components ($n=0$), are the only such possible discrete terms that can appear in a spectral density. Therefore, any on-shell state in a QFT must be attributed to the appearance of one such member of this class. In fact, in~\cite{Lowdon:2015fig} it was proven that an $n \geq 1$ generalised pole implies the existence of an on-shell state with \textit{vanishing} norm. Combining this with the well-known property that $\delta(s-m^{2})$ terms with coefficients of differing signs correspond to states with positive or negative inner products, this completes the classification of possible on-shell states in QFTs with an indefinite inner product. \\

\noindent
For (anti-)ghost fields several conventions exist in the literature regarding their Hermiticity assignment. Although seemingly innocuous, it turns out that these differing choices can potentially lead to inconsistencies in the BRST structure of the theory, including the violation of $S$-matrix unitarity~\cite{Nakanishi:1990qm}. By demanding that both $C^{a}$ and $\overbar{C}^{a}$ are Hermitian, and hence real, this avoids these problems. However, real fields inevitably imply that the spectral density $\rho_{C}^{ab}(s)$ in Eq.~\eqref{ghost_prop} must be complex, which raises questions about the relationship between $\rho_{C}^{ab}(s)$ and the state space structure of the theory. An overview of this relationship is detailed in Ref.~\cite{Nakanishi:1990qm}, where the authors demonstrate that the asymptotic ghost states are in fact constructed from the fields $C^{a}$ and $i\overbar{C}^{a}$. This ultimately implies that the spectral properties of states in the theory are instead captured by $i\rho_{C}^{ab}(s)$. In particular, given the appearance of a term $\mathcal{Z}^{ab}\delta(s-m^{2})$ in $i\rho_{C}^{ab}(s)$, the conditions $\mathcal{Z}^{ab}>0$ and $\mathcal{Z}^{ab}<0$ correspond to the existence of a state with positive or negative inner product, respectively.

\subsection{Euclidean spectral representation}
\label{Eucl_spec_rep}

In general, working with complex ghost fields is often more convenient since both the spectral density and Euclidean propagator are real. Although this can lead to theoretical issues, as outlined in Sec.~\ref{ghost_den}, in Landau gauge it has been proven that the real and complex ghost field formulations are equivalent~\cite{Alkofer:2000wg}. For the purposes of this study we will focus solely on lattice results in this gauge, and hence the distinction between real and complex ghost fields is no longer important. In most analyses of the Landau gauge ghost propagator, including the study~\cite{Duarte:2016iko} for which we analyse the lattice data in Sec.~\ref{lattice_fits}, the ghost fields are chosen to be complex, and so this convention will be assumed throughout the remainder of this paper. Although Eq.~\eqref{ghost_prop} holds for the real fields $C^{a}$ and $\overbar{C}^{a}$, the complex lattice fields $\eta^{a}$ and $\overbar{\eta}^{a}$ used in~\cite{Duarte:2016iko} are related as follows: $\eta^{a}=C^{a}$, $\overbar{\eta}^{a}=-i\overbar{C}^{a}$. By combining these relations with the spectral representation in Eq.~\eqref{ghost_prop}, one ultimately finds that the non-singular component of the Euclidean complex ghost propagator has the form    
\begin{align}
G_{E}^{ab}(p_{E}) =   \int_{0}^{\infty} \!\!ds \, \frac{ \rho_{\eta}^{ab}(s) }{p_{E}^{2}+s},  
\label{euclidean_rep}
\end{align}  
where $\rho_{\eta}^{ab}(s)=-i\rho_{C}^{ab}(s)$ and $p_{E}$ signifies the Euclidean momentum\footnote{For the remainder of this paper we will be dealing solely with imaginary-time quantities, and so for simplicity the Euclidean subscript on $p_{E}$ will be dropped.}. Since $\rho_{\eta}^{ab}(s)$ differs to the physical spectral density $i\rho_{C}^{ab}(s)$ by a sign, it immediately follows from the discussion in Sec.~\ref{ghost_den} and Eq.~\eqref{euclidean_rep} that the appearance of mass poles in $G_{E}^{ab}(p_{E})$ with coefficients $\mathcal{Z}^{ab}>0$ or $\mathcal{Z}^{ab}<0$ implies the existence of on-shell states with negative or positive inner products, respectively. For example, in the free theory the Euclidean propagator has the form $G_{E,\text{free}}^{ab} = \delta^{ab}/p^{2}$, which signifies the existence of a massless on-shell state with negative inner product, as expected.  

\subsection{Lattice ghost propagator}
\label{lattice_ghost}

In lattice formulations of Yang-Mills theory the Euclidean ghost propagator is defined by the inverse of the Faddeev-Popov operator $\mathcal{M}^{ab}=-\partial_{\mu}D_{\mu}^{ab}$ averaged over all field configurations
\begin{align}
G_{\text{lat}}^{ab}(p) =  \left\langle \sum_{x,y}   e^{-ip\cdot (x-y)} \left[\mathcal{M}^{-1} \right]^{ab}_{xy} \right\rangle,
\label{ghost_lattice}
\end{align}
where $D_{\mu}^{ab}$ is the lattice covariant derivative, and $p_{\mu}=\frac{2\pi}{L}k_{\mu}$ with $k_{\mu}=0, \dots, L-1$, and $L$ is the lattice size~\cite{Suman:1995zg}. In many studies it is often assumed that the propagator has the diagonal form $G_{\text{lat}}^{ab}(p)= \delta^{ab}G(p)$, as is the case in perturbation theory. Since on the lattice one computes the quantity $G(p)$, this is therefore defined 
\begin{align}
G(p)= \tfrac{1}{8}\sum_{a}G_{\text{lat}}^{aa}(p),
\label{ghost_trace}
\end{align}
where $a=1, \dots, 8$ because here we are interested in $\mathrm{SU}(3)$ Yang-Mills theory. However, the colour diagonality of $G_{\text{lat}}^{ab}(p)$ implicitly assumes that global colour symmetry is preserved non-perturbatively, which is \textit{a priori} unknown~\cite{Oehme:1979ai}. In order to keep maximal generality we will therefore not assume that this is the case. Since the spectral representation in Eq.~\eqref{euclidean_rep} exists independently of whether global colour symmetry is preserved or not, Eq.~\eqref{ghost_trace} implies that the lattice ghost propagator $G(p)$ has the following structure in the continuum limit  
\begin{align}
G(p) =   \int_{0}^{\infty} \!\!ds \, \frac{ \rho_{\eta}(s) }{p^{2}+s},  
\label{ghost_rep}
\end{align}
where $\rho_{\eta}(s):= \frac{1}{8}\sum_{a}\rho_{\eta}^{aa}(s)$. If colour symmetry is preserved then $\rho_{\eta}(s)$ defines a unique spectral density, whereas if the symmetry is broken $\rho_{\eta}(s)$ instead represents a colour-averaged quantity. In either case, it follows from Eq.~\ref{ghost_rep} that the analysis of lattice data for $G(p)$ amounts to probing the spectral properties of $\rho_{\eta}(s)$. \\ 

\noindent
Handling the appearance of Gribov copies is essential for performing a consistent lattice calculation in Yang-Mills theory. Due to the definition in Eq.~\eqref{ghost_lattice}, for the ghost propagator this amounts to ensuring that the Faddeev-Popov operator possesses a well-defined inverse. In BRST quantisation one also requires that the gauge fields satisfy a specific gauge-fixing condition, which in the case of Landau gauge is $\partial_{\mu}A_{\mu}^{a}=0$. It turns out that both of these conditions can be simultaneously satisfied by choosing gauge field configurations that minimise a specific functional of the fields $\mathcal{F}[A]$~\cite{Vandersickel:2012tz}. The resulting configurations are said to belong to the \textit{first Gribov region} (FGR). However, although the lattice path integral is well-defined in this region, there continue to exist Gribov copies corresponding to distinct minima of $\mathcal{F}[A]$. Since there is no unique way of handling these minima, different choices essentially result in different types of Landau gauge~\cite{Maas:2009se}. Minimal Landau gauge corresponds to taking an arbitrary choice of Gribov copy within the FGR for each field configuration~\cite{Vandersickel:2012tz}. For the purposes of this study we will focus on the ghost propagator in minimal Landau gauge, and in Sec.~\ref{lattice_fits} we will use lattice data in this gauge together with the analytic results outlined in this section in order to investigate the non-perturbative ghost spectrum.

\section{BRST symmetry and the ghost spectrum}
\label{brst_spec}

In BRST-quantised QCD the presence of BRST symmetry plays a central role in characterising the state-space structure of the theory, which in turn underpins many important characteristics including the unitarity of scattering for physical states, the characterisation of physical observables, and colour confinement~\cite{Nakanishi:1990qm}. Although it remains an open question as to whether BRST symmetry is realised beyond the perturbative regime, in what follows we will assume that this is indeed the case. By virtue of this symmetry, Kugo and Ojima proved that states with a negative inner product are necessarily absent from the physical space of states $\mathcal{V}_{\text{phys}}$, the so-called \textit{quartet mechanism}~\cite{Kugo:1979gm}. An important consequence of this argument is that the zero-norm states $\mathcal{V}_{0}$ in the theory are related to the full space of states $\mathcal{V}$ as follows:
\begin{align}
\mathcal{V}_{0} = Q_{B} \mathcal{V},
\label{zero_parent}
\end{align} 
where $Q_{B}$ is the BRST charge. In other words, any zero-norm state $|\Phi\rangle$ implies the existence of a partner state $|\widetilde{\Phi}\rangle \in \mathcal{V}$ such that $|\Phi\rangle = Q_{B}|\widetilde{\Phi}\rangle$. Since the physicality of states is defined by the condition $Q_{B}\mathcal{V}_{\text{phys}}=0$, and the quartet mechanism prevents states with negative inner product from appearing in $\mathcal{V}_{\text{phys}}$, for $|\Phi\rangle$ to be non trivial this therefore requires that the partner state $|\widetilde{\Phi}\rangle$ must have a negative inner product. Moreover, since $Q_{B}$ commutes with the operator $P^{2}$~\cite{Nakanishi:1990qm}, if $|\Phi\rangle$ is an on-shell state then $|\widetilde{\Phi}\rangle$ must necessarily have the same mass. Combining these results therefore implies
\vspace{1.5mm}
\begin{align}
&\textit{If a state $|\Phi\rangle$ in BRST-quantised QCD satisfies \hspace{0.1mm} $P^{2}|\Phi\rangle=m_{\Phi}^{2}|\Phi\rangle$, \ $\langle \Phi|\Phi\rangle=0$}  \nonumber \\
& \hspace{2mm}   \Longrightarrow \ \textit{There exists a corresponding state $|\widetilde{\Phi}\rangle$ with mass $m_{\Phi}$, but $\langle \widetilde{\Phi}|\widetilde{\Phi}\rangle<0$.}
\label{condition}
\end{align}
This conclusion is particularly relevant in light of the findings of Ref.~\cite{Li:2019hyv}. In this study, fits to infrared lattice data for the minimal Landau gauge gluon propagator were performed in order to test for different pole structures. By using ans\"{a}tze involving combinations of all possible classes of poles, evidence was found for the existence of a $(p_{E}^{2}+m_{A}^{2})^{-2}$ component, corresponding to a generalised term $\delta'(s-m_{A}^{2})$ in the gluon spectral density with
\begin{align}
m_{A} = 0.88  \substack{+0.09 \\ -0.06} \ \text{GeV}.
\end{align}
As discussed in Sec.~\ref{ghost_den}, the appearance of such a pole implies the existence of an on-shell zero-norm state $|\Psi_{A}\rangle$ with mass $m_{A}$, which can be interpreted as a gluonic excitation. The fact that $|\Psi_{A}\rangle$ has vanishing norm guarantees its confinement in QCD, since any such state has been shown to not contribute to physical scattering processes~\cite{Kugo:1979gm}. Applying the general condition in Eq.~\eqref{condition} to $|\Psi_{A}\rangle$ it immediately follows that if BRST symmetry is realised non-perturbatively, there must exist a corresponding partner state $|\widetilde{\Psi}_{A}\rangle$ also with mass $m_{A}$, but $\langle \widetilde{\Psi}_{A} |\widetilde{\Psi}_{A}\rangle<0$. An important question is whether this state could also be searched for using QCD propagator data. Initially one might have thought that this state could also contribute to the gluon propagator, and in contrast to $|\Psi_{A}\rangle$ would appear as an ordinary massive pole. However, since the partner state is related to $|\Psi_{A}\rangle$ via the action of the BRST charge $|\Psi_{A}\rangle = Q_{B} |\widetilde{\Psi}_{A}\rangle$, and $Q_{B}$ changes the ghost number of the state by one unit\footnote{See~\cite{Nakanishi:1990qm} and references within for more details regarding the ghost number and its corresponding generator.} this state, if it exists, would instead contribute to the ghost propagator $G(p)$. \\

\noindent
Using the analytic results of Sec.~\ref{spectral_ghost}, together with the conclusions drawn in this section, one is ultimately led to the following non-trivial test for the existence of non-perturbative BRST symmetry in minimal Landau gauge pure $\mathrm{SU}(3)$ Yang-Mills theory:
\begin{align}
\!\!\textit{If BRST symmetry exists \ $\Longrightarrow$ \ $G(p)$ must contain a component \hspace{0.1mm} $\frac{\mathcal{Z}}{p^{2}+m_{A}^{2}}$, where $\mathcal{Z}>0$.}
\label{condition2}
\end{align}
Taking the converse of this condition, this immediately implies that if the minimal Landau gauge ghost propagator \textit{does not} contain a massive pole at $p^{2}= -m_{A}^{2}$ with positive coefficient, it must be the case that BRST symmetry is not present on a non-perturbative level. This condition will be explored further in the next section using lattice ghost propagator data.
          
\newpage

\section{Lattice data fits}
\label{lattice_fits}

In this section we will outline the strategy we adopted to analyse lattice data for the minimal Landau gauge ghost propagator in pure $\mathrm{SU}(3)$ Yang-Mills theory. For this analysis we used the $\beta = 6.0$ data of Ref.~\cite{Duarte:2016iko} with $64^{4}$ and $80^{4}$ lattices, corresponding to physical volumes of $(6.50 \, \text{fm})^{4}$ and $(8.13 \, \text{fm})^{4}$ respectively, and a lattice spacing of $a=0.10 \, \text{fm}$. Full details of the gauge-fixing procedure and lattice setup can be found in Ref.~\cite{Duarte:2016iko}. Similarly to the study of the gluon propagator in Ref.~\cite{Li:2019hyv}, the goal of this analysis was to test for different pole structures in the infrared ghost lattice data. This is particularly interesting in light of the findings of Sec.~\ref{brst_spec}, since if these poles are present then they must dominate the behaviour of the propagator in the infrared, and hence searching for them constitutes a non-trivial test for the existence of on-shell states in the spectrum. In order to look for these components we analysed whether different propagator pole ans\"{a}tze could fit the lattice data up to some scale $p_{\text{max}}$. In particular, we considered both one and two term linear combinations, with the latter involving at least one standard non-generalised pole component
\begin{align}
&G_{i}(p) = \frac{z_{i}}{(p^{2} + m_{i}^{2})^{i+1}}, \label{ansatz1} \\
&G_{0j}(p) = \frac{z_{0}}{p^{2} + m_{0}^{2}}+\frac{Z_{j}}{(p^{2} + M_{j}^{2})^{j+1}},  \label{ansatz2}
\end{align}
where $i,j= 0,1,2$, and the masses $m_{0}$, $m_{i}$ and $M_{j}$ are real valued. The presence of different poles in the infrared region $p \in \left[0,p_{\text{max}}\right]$ means that the ghost spectral density $\rho_{\eta}(s)$ must contain discrete components. As as outlined in Sec.~\ref{ghost_den}, depending on the order of the pole and the sign of the coefficient, these components imply the existence of on-shell states with either positive, zero, or negative inner product. \\

\noindent   
Before discussing the more technical aspects of the fits, it is important to note that in this study we focused on the non-renormalised bare lattice data, similarly to the analysis of the gluon propagator in Ref.~\cite{Li:2019hyv}. Since renormalisation only results in an overall (positive) rescaling of the data~\cite{Duarte:2016iko,Dudal:2018cli}, the position and structure of any poles remains unchanged, and hence this information can be assessed from the bare data alone. The only difference is that the bare lattice propagator is a dimensionless quantity, and hence the coefficients $z_{i}$ and $Z_{j}$ must be dimensionful. Due to the difficulty of precisely accessing the systematic uncertainties in the lattice data we used three different choices of uncertainties in order to assess the robustness of the fits: statistical errors only, statistical errors plus a systematic shape uncertainty for momenta below $1 \,\text{GeV}$, and statistical errors plus a shape uncertainty for all values of momenta. In the second case, we simulated the shape uncertainty with the same empirical function as applied in Ref.~\cite{Li:2019hyv}, and in the final case we used a fourth-order polynomial in $p^{2}$ with deviations at the $2.5$\% level, which allowed for significant shape changes\footnote{This approach is identical to that used in the analysis of the gluon propagator in Ref.~\cite{Li:2019hyv}, except that the second uncertainty scenario only has a systematic shape uncertainty for momenta below $1 \,\text{GeV}$. We refer the reader to Ref.~\cite{Li:2019hyv} for more details.}. The logic behind the specific cut in the second uncertainty scenario follows from the findings of Ref.~\cite{Sternbeck:2012mf}, where the authors demonstrated that systematic effects, in particular the choice of Gribov copies, can lead to non-negligible shape changes in the ghost propagator for momenta below $1 \, \text{GeV}$. In all of these fitting scenarios the goodness of fit was assessed using a Chi-squared minimisation procedure, the specific details for which can be found in Ref.~\cite{Li:2019hyv}. The robustness of the fits were also tested by checking the sensitivity of the best-fit parameters to the momentum cutoff $p_{\text{max}}$. \\

\begin{figure}
\centering
\includegraphics[width=0.5\columnwidth]{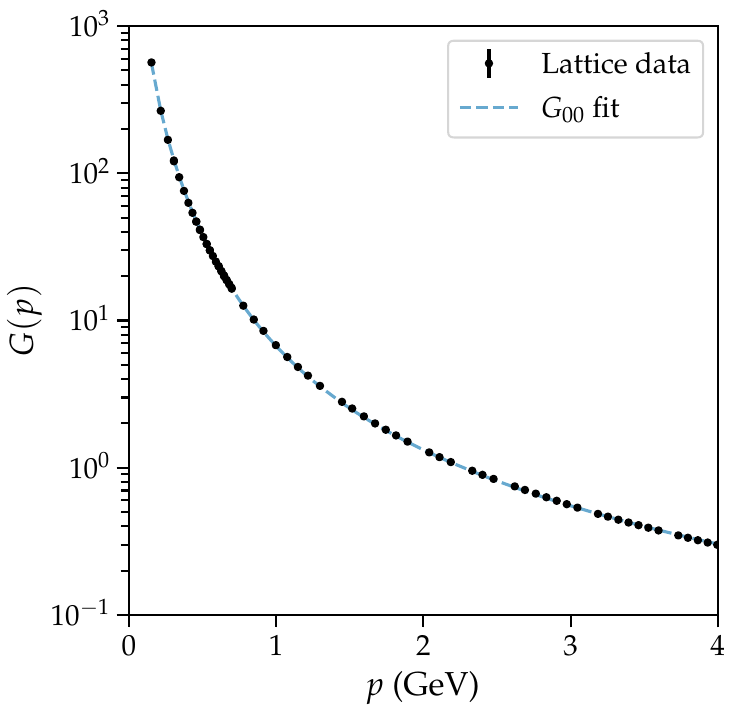}
\caption{Best-fit plot of the $G_{00}(p)$ ansatz together with the $80^{4}$ lattice data points. The statistical errors on the lattice data points are plotted, but too small to be seen.}
\label{fig:ansatze}
\end{figure}

\noindent
The $\chi^2$/d.o.f. values of the various fits to the $64^{4}$ and $80^{4}$ data are given in the Appendix. Comparing the different single-pole fits it was clear that the data strongly favoured an ordinary mass pole $G_{0}(p)$, with $m_{0}=0$. However, this component alone was not sufficient to describe the data over a range of momenta larger than $0.5 \, \text{GeV}$. Although the inclusion of any additional pole led to a significant improvement in the fits, $G_{00}(p)$ was the only ansatz that could provide a reliable fit in each of the systematic error scenarios, and whose best-fit parameter values remained stable across the different lattice volumes. When only statistical errors where included the highest quality fit for $G_{00}(p)$ was obtained from the $80^{4}$ data, resulting in the following best-fit parameter values:
\begin{align}
&z_{0} = 13.70  \pm 0.06 \ \text{GeV}^{2}, \quad m_{0} = 0.00 \substack{+0.02 \\ -0.00} \ \text{GeV},  \label{best_fit1} \\
&Z_{0} = -9.01  \substack{+0.25 \\ -0.17} \ \text{GeV}^{2}, \quad M_{0} = 0.54 \pm 0.03 \ \text{GeV},  \label{best_fit2}  
\end{align}
where the uncertainties indicate a 1$\sigma$ variation. This fit remained both convergent and physically consistent up to the cutoff $p_{\text{max}}=  4\, \text{GeV}$. In Fig.~\ref{fig:ansatze} the best-fit expression for $G_{00}(p)$ is plotted together with the $80^{4}$ lattice data points. 
\ \\  
 
\noindent
Due to the general structure in Eq.~\eqref{ghost_rep}, the compatibility of the lattice data with the parameter values in Eqs.~\eqref{best_fit1} and~\eqref{best_fit2} supports the conclusion that there exists a pair of on-shell ghost states in the spectrum: a massless state with negative inner product, and a positive-norm state with mass $M_{0} = 0.54 \ \text{GeV}$. Now whilst there are strong theoretical arguments for the appearance of a massless ghost state~\cite{Lowdon:2017gpp}, which is supported by several studies~\cite{Dudal:2019gvn, Horak:2021pfr, Siringo:2016jrc, Falcao:2020vyr}, the potential existence of an additional massive ghost state has to our knowledge not yet been considered in the literature. Although this massive state has positive norm, it also necessarily has a non-vanishing ghost number, and is hence ruled out from the physical spectrum of the theory~\cite{Kugo:1979gm}. Together with the detection of this additional ghost state, another important finding from this analysis is that the ghost propagator lattice data is \textit{not} compatible with the existence of a component $\mathcal{Z}(p^{2}+m_{A}^{2})^{-1}$ where $\mathcal{Z}>0$, and hence a negative-norm state with mass $m_{A}$. Taking the contrapositive of the condition in Eq.~\eqref{condition2}, one is ultimately led to the conclusion that for pure $\mathrm{SU}(3)$ Yang-Mills theory 
\begin{align}
\textit{BRST symmetry is not realised non-perturbatively in minimal Landau gauge.}
\label{statement}  
\end{align}
A similar conclusion has also been reached in other studies. In Refs.~\cite{Fischer:2008uz,vonSmekal:2008ws} the authors argue that the existence of Gribov copies in minimal Landau gauge is sufficient to imply that global BRST symmetry must be broken, and in Ref.~\cite{Burgio:2009xp} the ill-definedness of the BRST charge is linked to the Gribov-Zwanziger mechanism. More recently, in Ref.~\cite{Maas:2019ggf} lattice data for $\mathrm{SU}(2)$ Yang-Mills theory was used to demonstrate that the minimal Landau gauge ghost propagator is not consistent with the continuum Dyson-Schwinger equation (DSE). Since the DSE structure is predicated on the existence of BRST symmetry, this therefore implies an analogous conclusion to Eq.~\ref{statement}. An appealing characteristic of the argument leading to Eq.~\ref{statement} is that it derives from the fact that BRST symmetry imposes specific spectral constraints, and hence the demonstration that one of these constraints is violated, in this case the absence of a corresponding ghost partner state with mass $m_{A}$, is sufficient to imply that the symmetry cannot be present non-perturbatively. Moreover, the theoretical assumptions underpinning this argument are broad, relying only on the existence of a K\"{a}ll\'{e}n-Lehmann representation, the structure of which means that the on-shell states in the theory can be detected by analysing the infrared behaviour of the corresponding field propagators.

\section{Conclusions}
\label{concl}

The presence of BRST symmetry in BRST-quantised Yang-Mills theory has many implications for the non-perturbative structure of the theory. In particular, in QCD this symmetry implies a non-trivial connection between the ghost and gluon spectra, namely that on-shell zero-norm gluonic states must possess a corresponding ghost partner state with identical mass, but negative inner product. Since the discrete spectrum of the theory can be inferred from the infrared behaviour of the field propagators, by analysing lattice data for the ghost and gluon propagators this provides a non-trivial test of whether BRST symmetry is realised non-perturbatively in lattice formulations of the theory. Using lattice data for the minimal Landau gauge ghost propagator in pure $\mathrm{SU}(3)$ Yang-Mills theory, we find evidence for the existence of a pair of on-shell ghost states in the spectrum: a massless state with negative inner product, and a positive-norm state with mass $M_{0} = 0.54 \ \text{GeV}$. In doing so, we rule out the existence of a potential ghost partner state associated with the zero-norm gluonic state established in Ref.~\cite{Li:2019hyv}, and hence conclude that BRST symmetry cannot be realised non-perturbatively in minimal Landau gauge. Ultimately, this implies that continuum and current lattice formulations of Yang-Mills theory in Landau gauge represent two distinct realisations of the theory.

\section*{Acknowledgements}
The authors would like to thank Axel Maas and Jan Pawlowski for useful discussions and input. S.~W.~L is supported by the Fermi Research Alliance, LLC under Contract No.~DE-AC02-07CH11359 with the U.S. Department of Energy, Office of Science, Office of High Energy Physics, and P.~L. by the Deutsche Forschungsgemeinschaft (DFG, German Research Foundation) through the Collaborative Research Center CRC-TR 211 ``Strong-interaction matter under extreme conditions'' Project No.~315477589-TRR 211. P.~J.~S. acknowledges support from FCT under the contract CEECIND/00488/2017, and both O.~O. and P.~J.~S. acknowledge support by national funds from FCT -- Funda\c{c}\~{a}o para a Ci\^{e}ncia e a Tecnologia, I.P., within the projects UIDB/04564/2020 and UIDP/04564/2020, and also from CERN/FIS-COM/0029/2017. This work was granted access to the HPC resources of the PDC Center for High Performance Computing at the KTH Royal Institute of Technology, Sweden, made available within the Distributed European Computing Initiative by the PRACE-2IP, receiving funding from the European Community's Seventh Framework Programme (FP7/2007-2013) under grand agreement No.~RI-283493. The use of Lindgren has been provided under DECI-9 project COIMBRALATT. We acknowledge that the results of this research have been achieved using the PRACE-3IP project (FP7 RI312763) resource Sisu based in Finland at CSC. The use of Sisu has been provided under DECI-12 project COIMBRALATT2. We also acknowledge the Laboratory for Advanced Computing at the University of Coimbra (\nolinkurl{http://www.uc.pt/lca}) for providing access to the HPC resource Navigator.

\appendix
\section{Goodness-of-fit results}

The goodness-of-fit results for the $G_{i}(p)$ and $G_{0j}(p)$ ans\"{a}tze with the $64^{4}$ and $80^{4}$ lattice data of Ref.~\cite{Duarte:2016iko} are found in Tables~\ref{tab:chi_square_64} and~\ref{tab:chi_square_80}, respectively. In each fit a value of $p_{\text{max}} =1 \, \text{GeV}$ was initially chosen to determine whether a convergent and physically consistent ($p_{\text{max}}> m_{0}, m_{i}, M_{j}$) fit could be achieved. If not, the value of $p_{\text{max}}$ was lowered or raised until these conditions were satisfied. The values of $p_{\text{max}}$ in the tables reflect this final choice of cutoff. The three columns represent the fits performed using the three different systematic error scenarios outlined in Sec.~\ref{lattice_fits}.
 
\ \\
\begin{table}[!ht]
\small
    \begin{center}
	\begin{tabular}{|c |c| c| c |c|}
	\hline
	 & \textbf{Stat. only} & \textbf{Stat. + Shape $p< 1 \, \text{GeV}$} & \textbf{Stat. + Shape all $p$} \\
	 & $\chi_{1}^{2}$/d.o.f.  ($p_{\text{max}}$) & $\chi_{2}^{2}$/d.o.f.  ($p_{\text{max}}$) & $\chi_{3}^{2}$/d.o.f. ($p_{\text{max}}$) \\
	\hline
	$G_0(p)$ & $>80$ \ (0.5) & $>20$ \ (0.5) & $>60$ \ (0.5) \\  
	\hline
	$G_1(p)$ & $>20$ \ (0.5) & $>40$ \ (0.5) & $>20$ \ (0.5)\\
	\hline
	$G_2(p)$ & $>60$ \ (0.5) & $>70$ \ (0.5) & $>70$ \ (0.5) \\
	\hline 
	$G_{00}(p)$ & 0.94  \ (4.0) & 0.82  \ (4.0) &  0.4 \ (4.0) \\  
	\hline
	$G_{01}(p)$ & 1.2  \ (1.2) &  0.9 \ (1.0) & 1.0  \ (1.0) \\
	\hline
	$G_{02}(p)$ & 6.7  \ (1.6) & 2.0 \ (1.0) & 1.7 \ (1.0) \\
	\hline    
	\end{tabular}
    \caption{Chi-squared fit results for $G_{i}(p)$ and $G_{0j}(p)$ under the different systematic error scenarios with the $64^{4}$ lattice data.}
    \label{tab:chi_square_64}
    \end{center}    
\end{table}

\begin{table}[!ht]
\small
    \begin{center}
	\begin{tabular}{|c |c| c| c |c|}
	\hline
	 & \textbf{Stat. only} & \textbf{Stat. + Shape $p< 1 \, \text{GeV}$} & \textbf{Stat. + Shape all $p$} \\
	 & $\chi_{1}^{2}$/d.o.f.  ($p_{\text{max}}$) & $\chi_{2}^{2}$/d.o.f.  ($p_{\text{max}}$) & $\chi_{3}^{2}$/d.o.f. ($p_{\text{max}}$) \\
	\hline
	$G_0(p)$ & $>300$ \ (0.5) & $>10$ \ (0.5) & $>20$ \ (0.5) \\  
	\hline
	$G_1(p)$ & $>200$ \ (0.5) & $>100$ \ (0.5) & $>20$ \ (0.5)\\
	\hline
	$G_2(p)$ & $>500$ \ (0.5) & $>200$ \ (0.5) & $>50$ \ (0.5) \\
	\hline 
	$G_{00}(p)$ & 1.2 \ (4.0) & 1.2 \ (4.0) & 0.9 \ (4.0) \\  
	\hline
	$G_{01}(p)$ & 1.6 \ (0.5) & 1.6 \ (0.5) & 1.8 \ (0.7) \\
	\hline
	$G_{02}(p)$ & 3.4 \ (0.5) & 3.2 \ (0.5) & 1.8 \ (0.5) \\
	\hline    
	\end{tabular}
    \caption{Chi-squared fit results for $G_{i}(p)$ and $G_{0j}(p)$ under the different systematic error scenarios with the $80^{4}$ lattice data.}
    \label{tab:chi_square_80}
    \end{center}    
\end{table}

\bibliographystyle{JHEP}

\bibliography{refs}

\end{document}